\documentclass[12pt,aps,prd,showpacs,superscriptaddress,nofootinbib]{revtex4-2}
\pdfoutput=1
\usepackage{braket,bm}
\usepackage[pass]{geometry} 
\usepackage{color,graphicx}
\usepackage{cancel}
\usepackage{caption}
\usepackage{subcaption}
\usepackage[colorlinks,citecolor=blue,linkcolor=black]{hyperref} %if we want hyperlinked references
\usepackage{bm}        
\usepackage{amssymb}   
\usepackage{mathrsfs,mathtools, wasysym}
\usepackage[section]{placeins} %keeps figures in the right section
\DeclareMathAlphabet{\mathpzc}{OT1}{pzc}{m}{it}

\usepackage{ushort}
\usepackage[normalem]{ulem}

\newcommand{\D}{\mathrm{d}}

\begin{document}

%%%%%%%%%%%%%%%%%%%%%%%%%%%%%%%%%%%%%%%%%%%%%%%%%%%%%%%%%%%%%%%%%%%%%%
%\preprint{MAN/HEP/2024/XXX}
\title{A new study of the Unruh effect}
\author{Robert Dickinson}
%\email{robert.dickinson-2@manchester.ac.uk}
\author{Jeff Forshaw}
%\email{jeff.forshaw@manchester.ac.uk}
\author{Ross Jenkinson\footnote{Author to whom any correspondence should be addressed.}}
%\email{ross.jenkinson@manchester.ac.uk}
\affiliation{Department of Physics and Astronomy,
  University of Manchester,
  Manchester M13 9PL,
  United Kingdom}
\author{Peter Millington}
%\email{p.millington@nottingham.ac.uk}
\affiliation{Department of Physics and Astronomy,
  University of Manchester,
  Manchester M13 9PL,
  United Kingdom}

\date{\today}

\begin{abstract}
We revisit the Unruh effect within a general framework based on direct, probability-level calculations. We rederive the transition rate of a uniformly accelerating Unruh-DeWitt monopole detector coupled to a massive scalar field, from both the perspective of an inertial (Minkowski) observer and an accelerating (Rindler) observer. We show that, for a measurement at a finite time after the initial state is prepared, the two perspectives give the same transition rate. We confirm that an inertial detector in a thermal bath of Minkowski particles responds differently to the accelerated detector (which perceives a thermal bath of Rindler particles), except in the case of a massless field where there is agreement at all times. Finally, new numerical results for the transition rate are presented and explained, highlighting the transient effects caused by forcing the field to initially be in the Minkowski vacuum state.
\end{abstract}


\maketitle

\section{Introduction}

The trajectories of uniformly accelerating observers (\textit{Rindler observers}) are restricted to a region of Minkowski spacetime (the \textit{Rindler wedge}), and they are causally disconnected from another region of Minkowski spacetime (the opposite Rindler wedge). The mode expansion of a quantum field employed by a Rindler observer is different to that employed by a Minkowski observer. Thus, accelerated and inertial observers may disagree on the particle content of a field. Remarkably, a Rindler observer would associate a thermal bath of Rindler particles to the no-particle (vacuum) Minkowski state. This is the Unruh effect~\cite{Unruh:1976db, Fulling:1972md, Davies:1974th, Unruh:1977}.

The Unruh effect is a direct mathematical consequence of quantum field theory. To probe the physics of the Unruh effect, localized particle detector models were developed and applied for a uniformly accelerating path~\cite{Unruh:1976db, Unruh:1977, DeWitt:1975ys}. The conclusion is that the detector's non-zero response rate per unit proper time along the detector's trajectory as measured by a Minkowski observer (who, using inertial measuring apparatus, otherwise experiences a vacuum) is identical to that measured by a Rindler observer (who, using rigidly accelerating apparatus, experiences a thermal heat bath).
The effect can be understood as a consequence of the presence of a horizon, which appears between the Rindler wedge and the rest of the universe.  Therefore, similar methods to those used to study the Unruh effect can be used to study horizons in curved spacetimes~\cite{Sewell1982, KayWald1991}, reproducing the thermal properties of black holes~\cite{Bekenstein1973, Hawking1974Explosions, Hawking:1974sw, GibbonsPerry1978} and de Sitter space~\cite{GibbonsHawking1977}. The mathematical relationship between Minkowski and Rindler coordinates is very similar to that between Schwarzschild and Kruskal coordinates for black holes. A significant difference is that Hawking radiation is detectable at infinity, since Schwarzschild coordinates become inertial at large distances. Close to the horizon, an observer at a fixed radial position would detect thermal effects that a free-falling observer would not~\cite{Unruh:1976db, Unruh1977b, Fulling1977}, and this can be attributed to the acceleration required to maintain constant radial position. There are also key similarities between the rotational Unruh effect and rotating black holes~\cite{Starobinsky1973, Unruh1974}. Consequently, the Unruh effect offers an excellent avenue into understanding important features of quantum field theory that are also relevant to Hawking radiation and black holes.

Such key features of quantum field theory are incorporated further in the field of relativistic quantum information, which focuses on the relationship between relativistic quantum field theories and quantum information~\cite{Mann_2012, Tjoa_2024}. This has resulted in important applications of the Unruh effect and the Unruh-DeWitt detector, such as entanglement harvesting~\cite{PhysRevLett.95.120404, VALENTINI1991321, Salton_2015, PhysRevD.105.085012, PhysRevD.104.025001, Tjoa2020, PhysRevD.106.076002, PhysRevA.82.042332}, entanglement degradation~\cite{PhysRevD.51.1716, PhysRevD.68.085006, PhysRevD.82.064006, PhysRevA.80.042318, PhysRevA.81.032320}, corrections to quantum teleportation fidelity~\cite{Alsing:2003cy, PhysRevA.74.032326, PhysRevLett.91.180404}, quantum energy teleportation~\cite{PhysRevA.89.012311}, curvature measurement~\cite{Martin-Martinez_2012, PhysRevD.89.024013}, and avoiding difficulties with field measurements~\cite{Sorkin:1993gg, Benincasa_2014, PhysRevD.103.025017,
Dragan_2013}.

In Section \ref{sec:accel}, the transition rate of a uniformly accelerated Unruh-DeWitt monopole detector is calculated using a general, probabilistic method. This method has been applied to other source-detector setups, such as the Fermi two-atom problem and scattering experiments~\cite{Dickinson:2013lsa, Dickinson:2016oiy, Dickinson:2017uit, Dickinson:2017gtm}, where the ease within which it can sum inclusively over final states has been proven to make causality manifest. The transition rate is calculated for a measurement a finite time after preparing the initial state. Specifying the field to initially be in the Minkowski vacuum state causes transients, which decay as the measurement time increases, and these transients are investigated. In Section \ref{sec:rindlerbath}, the same transition rate is calculated from the perspective of an accelerating Rindler observer. The result is the same, including the finite-time transient effects. The corresponding transition rate for an inertial detector in a bath of Minkowski particles is calculated in Section \ref{sec:minkowskibath}, and it is shown that this is different, except for the massless case. Section \ref{sec:numericalresults} presents our numerical results, and Section \ref{sec:conc} concludes.

Throughout this paper, we adopt natural units $c = \hbar = k_{\rm{B}} = 1$ and the `mostly-minus' metric signature $(+ - - -)$.

\section{Excitation rate of an accelerated detector}\label{sec:accel}

We consider a point-like `atom', $D$, which plays the role of a two-state, Unruh-DeWitt detector. The atom interacts with a neutral scalar field $\phi(\mathbf{x},t)$, of mass $m$, where $\mathbf{x}$ and $t$ are co-ordinates in an inertial frame, and it is accelerated with a constant proper acceleration, $\alpha$, such that its position is given by
\begin{align}
\label{eq:trajectory}
  \mathbf{x}^D = \bigg( \frac{1}{\alpha} {\rm cosh}\, \alpha \tau \,,\, 0 \,,\, 0 \bigg) = \bigg( \frac{1}{\alpha} \sqrt{1 + \alpha^2 t^2} \,,\, 0 \,,\, 0 \bigg) \, ,
\end{align}
and the proper time of the atom is
\begin{equation} \label{eq:propertime}
   \tau =\int_0^t\frac{{\rm d}t'}{\gamma(t')}=\int_0^{t}\frac{{\rm d}t'}{\sqrt{1+\alpha^2t'^2}}=\frac{1}{\alpha}\,{\rm arcsinh}\,\alpha t\,.
\end{equation}
The system is described by states living in a product of the Hilbert spaces of the atom and the field: $\mathscr{H}=\mathscr{H}^D\times\mathscr{H}^{\phi}$. For the Hamiltonian, we take $H(t)=H_0(t)+H_{\rm int}(t)$, where $H_0(t)=H_0^D(t)+H_0^{\phi}(t)$. Under the free part of the Hamiltonian, $H_0$, the atom has a complete set of states $\{\ket{1^D},\ket{2^D}\}$ (one ground state $\ket{1^D}$ and one excited state $\ket{2^D}$), with $H_0^D\ket{n^D}=\omega_n \ket{n^D}$, $n=1,2$. In the inertial frame, we assume that the interaction-picture Hamiltonian is given by
\begin{subequations}
\begin{align}
H_0 \ &=\  \sum_{n\,=\,1}^2\gamma^{-1}(t)\,\omega_n \ket{n^D}\bra{n^D}\:+\: \int \D^3\mathbf{x}\; \Big( \tfrac{1}{2}(\partial_t\phi)^2\:+\: \tfrac{1}{2}(\nabla\phi)^2\: +\: \tfrac{1}{2}m^2\phi^2 \Big)\,,\\
H_{\mathrm{int}} \ &=\ M^D(t)\,\phi(\mathbf{x}^D,t)\,, \label{eq:h1ex}
\end{align}
\end{subequations}
where $M^D(t) \equiv \gamma^{-1}(t)\sum_{mn}\mu_{mn} \, e^{i(\omega_m-\omega_n)\tau} \,\ket{m^D}\bra{n^D}$ represents a monopole interaction. For future reference, we define
$\mu\equiv \mu_{12}=\mu_{21}^{*}$ and $\omega\equiv \omega_2 - \omega_1$. 
We will also assume that $\mu_{nn}=0\;\forall\,n$, so that the interaction always involves transitions between the states. 

Suppose that the system is initially ($t=0$) described by a density matrix $\rho_0$ and that the measurement outcome is described by an effect operator $E$. In general, $E$ is an element of a Positive Operator-Valued Measure, and it may be written as a sum over products of Hermitian operators:
\begin{align}
E \ &=\  \sum_\kappa E^D_{(\kappa)} \otimes E^{\phi}_{(\kappa)} ~.
\end{align}
The superscripts $D$ and $\phi$ denote the Hilbert space in which the operators act and $\kappa$ denotes different configurations of final states. The probability of the measurement outcome, $\mathbb{P}$, is then given by 
\begin{equation}
\mathbb{P} \ =\  \mathrm{Tr} (E  \rho_t)\,, \label{eq:prob}
\end{equation}
where
\begin{equation}
\rho_t \ \equiv\  U_{t,0}\,\rho_0 \,U^\dag_{t,0}
\end{equation}
is the density operator at time $t$ and
\begin{equation}
U_{t,0} \ =\  \mathrm{T}\exp\bigg(\,\frac{1}{i}\!\int_0^{t} \! \D t'\; H_{\mathrm{int}}(t') \bigg)
\end{equation}
is the unitary evolution operator ($\mathrm{T}$ indicates time ordering).

We consider the case where the initial density operator is $\rho_0=\ket{1^D,0_{\phi}^{M}}\bra{1^D,0_{\phi}^{M}}$, in which $\ket{0^M_\phi}$ denotes the Minkowski vacuum state, and the effect operator is $E=\ket{2^D}\bra{2^D}\otimes \mathbb{I}_{\phi}$. This effect operator describes a set of final states in which the atom is excited and the final state of the field is anything at all. Fixing the field in the Minkowski vacuum state, $\ket{0_\phi^M}$ at an instant in time $(t=0)$ is somewhat arbitrary and will result in transient effects. 

We are interested in the excitation rate of the atom, $\varGamma(1\to 2)$. The master equation for the probability of finding the detector in the excited state is given by
\begin{equation}
\frac{{\rm d}\mathbb{P}(2;t)}{{\rm d}t}\ =\ \varGamma(1\to 2)\mathbb{P}(1;t)\:-\:\varGamma(2\to 1)\mathbb{P}(2;t)\,,
\end{equation}
where
\begin{equation}
\varGamma(1\to 2)\  =\ \frac{{\rm d}\mathbb{P}(2;t)}{\mathrm{d} t}\Big(1\:+\:\mathcal{O}(|\mu|^2)\Big)\,,
\end{equation}
and
\begin{equation}
	\mathbb{P}(2;t) \equiv \bra{1^D,0_{\phi}^{M}} U^\dag_{t,0} E \, U_{t,0} \ket{1^D,0_{\phi}^{M}}\,.
\end{equation}

Following \cite{Dickinson:2016oiy}, we use a generalization of the Baker-Campbell-Hausdorff lemma to commute the operator $E$ through the time-evolution operator, which gives  
\begin{align}
\mathbb{P}(2;t) 	\ &=\  \sum_{j\,=\,0}^\infty \int_0^{t} \D t_1 \D t_2\ldots \D t_j \;\Theta_{12...j} \bra{1^D,0_{\phi}^{M}}\mathcal{F}_j\ket{1^D,0_{\phi}^{M}}\,, \label{eq:fsum}
\end{align}
where
\begin{align}
\mathcal{F}_0 \ &=\ E \,, \nonumber\\
\mathcal{F}_j \ &=\ \tfrac{1}{i}\Big[ \mathcal{F}_{j-1}, H_\mathrm{int}(t_j) \Big]\,,
\label{eq:effs}
\end{align}
and $\Theta_{ijk\ldots} \equiv 1$ if $t_i>t_j>t_k\ldots$ and zero otherwise.
Using the notation $\phi_j^D \equiv \phi(\mathbf{x}_j^D,t_j)$, $M^D_j \equiv M^D(t_j)$ and $\mathbf{x}_j^D\equiv \mathbf{x}^D(t_j)$, we may write
\begin{align}
\mathcal{F}_j \ &=\ \tfrac{1}{i}\Big[ \mathcal{F}_{j-1}\,,\, M^D_j\phi^D_j \Big]\,.
\end{align}
Eq.~\eqref{eq:fsum} includes contributions from all perturbative orders. 

Thus far, we have introduced a general method for calculating observable probabilities that is based on effect operators and the evolution of the initial density matrix. This approach has the advantage that it can be used to treat both pure and mixed states, as well as exclusive, inclusive and semi-inclusive observables~\cite{Dickinson:2017uit}, all on equal footing. Moreover, it is an approach that has been shown to make physical principles, such as causality~\cite{Dickinson:2013lsa}, manifest. In what follows, we show how this probability-level approach can be used to treat the initial time-dependent response of the Unruh-DeWitt detector and recover known results in the late-time limit. This is with a view to future applications to problems that may be less tractable at the amplitude level.

Proceeding to expand in $\mu$, the two lowest-order contributions are
\begin{align}
\mathcal{F}_1\ &=\ i\gamma^{-1}(t_1)\phi_1^D\braket{2|2}\Big(\mu\,e^{-i\omega \tau_1}\ket{1}\!\bra{2}\:-\:\mu^*e^{i\omega \tau_1}\ket{2}\!\bra{1}\Big)\,,\\
\mathcal{F}_2\ &=\ \gamma^{-1}(t_1)\gamma^{-1}(t_2)\braket{2|2}|\mu|^2\Big[\mathbb{I}_{\phi}\,\Delta_{12}\sin(\omega\tau_{12})\Big(\braket{2|2}\ket{1}\!\bra{1}\:+\:\braket{1|1}\ket{2}\!\bra{2}\Big)\nonumber\\&\qquad+\:\{\phi_1^D,\phi_2^D\}\cos(\omega\tau_{12})\Big(\braket{2|2}\ket{1}\!\bra{1}\:-\:\braket{1|1}\ket{2}\!\bra{2}\Big)\Big]\,,\label{eq:F2}
\end{align}
where
\begin{equation} \label{eq:paulijordan}
\Delta_{12}\ \equiv\ \frac{1}{i}\,\braket{0_{\phi}^{M}|[\phi_1^D,\phi_2^D]|0_{\phi}^{M}}
\end{equation}
is the Pauli-Jordan function of the $\phi$ field (evaluated at points on the detector's path) and
\begin{equation}
\tau_{12}\equiv \tau_1-\tau_2  > 0\,.
\end{equation}
The lowest-order, non-vanishing contribution to $\mathbb{P}$ arises from $\mathcal{F}_2$ in Eq.\eqref{eq:F2}, and is
\begin{align}
\label{eq:Pmid}
\mathbb{P}(2;t) 	\ &=\ |\mu|^2\int_0^{t} \frac{\D t_1}{\gamma(t_1)} \int_{0}^{t_1}\frac{\D t_2}{\gamma(t_2)}\;\Big[\Delta^{R}_{12}\sin(\omega \tau_{12})\:+\:\Delta^{H}_{12}\cos(\omega \tau_{12})\Big]\nonumber\\
&=\ |\mu|^2\int_0^{\tau} \D \tau_1\int_{0}^{\tau_1}\D \tau_2\;\Big[\Delta^{R}_{12}\sin(\omega \tau_{12})\:+\:\Delta^{H}_{12}\cos(\omega \tau_{12})\Big]\,,
\end{align}
where, due to the time-ordering, $\Delta_{12}$ has become the retarded propagator 
\begin{equation}
\Delta^{R}_{12}\ \equiv\ \Theta_{12} \,\Delta_{12} \,,
\end{equation}
and
\begin{equation}
\Delta^{H}_{12}\ \equiv\ \braket{0_{\phi}^{M}|\{\phi_1^D,\phi_2^D\}|0_{\phi}^{M}}
\end{equation}
is the Hadamard propagator of the $\phi$ field (evaluated at points on the detector's path). Eq.~\eqref{eq:Pmid} is equal to
\begin{equation}
    \mathbb{P}(2;t) =\ |\mu|^2\int_0^{\tau} \D \tau_1\int_{0}^{\tau}\D \tau_2\;e^{-i\omega(\tau_1 - \tau_2)} \braket{0_{\phi}^{M}|\phi_1^D \,\phi_2^D|0_{\phi}^{M}}\,,
\end{equation}
as calculated by DeWitt~\cite{DeWitt:1980hx} and seen in many papers thereafter (albeit with integration limits $-\infty < \tau_1 , \tau_2 < \infty$). Since the commutator of interaction-picture fields is proportional to the identity operator, the free retarded propagator $\Delta^R$ does not depend on the initial state{\footnote{This is no longer the case when the retarded propagator is dressed with self-energy corrections.}. Therefore, writing the rate in terms of $\Delta^R$ and $\Delta^H$ separates it into a term which does not depend on the initial state and a term which does.

For time-like intervals, $\Delta^{R}_{12}$ and $\Delta^{H}_{12}$ are given by~\cite{Dickinson:2016oiy}\footnote{Note that there is a sign error on the $Y_1(z_{ij})$ term in equation (A11) in \cite{Dickinson:2016oiy} (see~\cite{Greiner:1996zu}).},
\begin{equation}
\label{eq:timelikepropagators}
\Delta^{R}_{12}\ =\ \frac{m^2}{4\pi}\frac{J_1(ms^\alpha_{12})}{ms^\alpha_{12}} - \frac{\delta((s^\alpha_{12})^2)}{2\pi} 
\quad \text{and} \quad
\Delta^{H}_{12}\ =\ \frac{m^2}{4\pi}\frac{Y_1(ms^\alpha_{12})}{ms^\alpha_{12}} \, ,
\end{equation}
where $J_1$ and $Y_1$ are Bessel functions of the first and second kind, and
\begin{align}
s^\alpha_{12} \ &\equiv  \sqrt{(x^{\mu}_1 - x^{\mu}_2)^2} =\ \sqrt{(t_1-t_2)^2 - \frac{1}{\alpha^2} \left( \sqrt{1+\alpha^2t_1^2} - \sqrt{1+\alpha^2t_2^2} \right)^2} \ =\ \frac{2}{\alpha}\sinh\frac{\alpha\tau_{12}}{2} \,,
\end{align}
for the trajectory given in Eq.~\eqref{eq:trajectory}. The delta function in Eq.~\eqref{eq:timelikepropagators} only has support at $\tau_{12} = 0$ and will not contribute further (the coefficient of $\Delta^{R}_{12}$ will always vanish at $\tau_{12} = 0$).
Eq.~\eqref{eq:Pmid} thus becomes
\begin{align}
\mathbb{P}(2;t) =\ |\mu|^2\int_0^{\tau} \D \tau_1 \int_{0}^{\tau_1}\D \tau_2 \;\frac{m^2}{4\pi} \;\left[ \frac{J_1(ms^\alpha_{12})}{ms^\alpha_{12}}\sin(\omega\tau_{12}) + \frac{Y_1(ms^\alpha_{12})}{ms^\alpha_{12}}\cos(\omega\tau_{12}) \right]\,.
\end{align}
This expression is divergent for $\tau_{12} = 0$, since $Y_1(x) \rightarrow - \infty$ as $x \rightarrow 0$. This is because the two-point correlation function evaluated at a point is infinite, and the divergence is not present if one considers a detector of finite spatial extent~\cite{Unruh:1976db, DeWitt:1980hx, GroveOttewill1983, Takagi1986}. The expression can be regularised by considering a spatial profile~\cite{Schlicht:2003iy, Louko:2006zv, Louko:2007mu}, but it would remain the case that a measurement at $\tau = 0$ would be divergent. Changing variables and introducing a lower limit on the integral to cut off the divergent part, the transition probability becomes
\begin{align}
\mathbb{P}(2;t) 	\ &=\ |\mu|^2\int_{1/\Lambda}^{\tau} \D \tau_1 \int_{1/\Lambda}^{\tau_1}\D \tau_{12} \;\frac{m^2}{4\pi} \;\left[ \frac{J_1(ms^\alpha_{12})}{ms^\alpha_{12}}\sin(\omega\tau_{12}) + \frac{Y_1(ms^\alpha_{12})}{ms^\alpha_{12}}\cos(\omega\tau_{12}) \right] \,.
\end{align}
The lowest-order transition rate is then
\begin{align}
\label{eq:accelrate}
\frac{\partial \mathbb{P}}{\partial \tau} \ &=\ |\mu|^2 \int_{1/\Lambda}^{\tau}\D \tau_{12}\;\Big[\Delta^{R}_{12}\sin(\omega \tau_{12})\:+\:\Delta^{H}_{12}\cos(\omega \tau_{12})\Big] \\
%&=\ \frac{m^2|\mu|^2}{4\pi} \int^\tau_{1/\Lambda} \D \tau_2 \;\bigg[ \frac{J_1(ms^\alpha_{\tau 2})}{ms^\alpha_{\tau 2}}\sin\omega(\tau\!-\!\tau_2) + \frac{Y_1(ms^\alpha_{\tau 2})}{ms^\alpha_{\tau 2}}\cos\omega(\tau\!-\!\tau_2) \bigg] \\
&=\ \frac{m^2|\mu|^2}{4\pi} \int^\tau_{1/\Lambda} \D \tau_{12} \;\bigg[ \frac{J_1(ms^\alpha_{12})}{ms^\alpha_{12}}\sin\omega\tau_{12} + \frac{Y_1(ms^\alpha_{12})}{ms^\alpha_{12}}\cos\omega\tau_{12} \bigg] \,. \label{eq:besselrate}
\end{align}
%where $s' = \frac{2}{\alpha}\sinh\frac{\alpha \tau'}{2}$.
Consider this transition rate with the inertial ($\alpha=0$) case subtracted,
\begin{align}
\label{eq:besselratesubtracted}
\frac{\partial \mathbb{P}}{\partial \tau} - \left.\frac{\partial \mathbb{P}}{\partial \tau}\right|_{\alpha = 0} %&=\ \frac{m^2|\mu|^2}{4\pi} \int^\tau_{1/\Lambda} \D \tau_2 \;\bigg[ \frac{J_1(ms^\alpha_{\tau 2})}{ms^\alpha_{\tau 2}}\sin\omega(\tau\!-\!\tau_2) + \frac{Y_1(ms^\alpha_{\tau 2})}{ms^\alpha_{\tau 2}}\cos\omega(\tau\!-\!\tau_2) \bigg] \\
&=\ \frac{m^2|\mu|^2}{4\pi} \int^\tau_{1/\Lambda} \D \tau_{12} \;\bigg[ \bigg( \frac{J_1(ms^\alpha_{12})}{ms^\alpha_{12}} - \frac{J_1(m\tau_{12})}{m\tau_{12}} \bigg) \sin\omega\tau_{12} \nonumber \\
& \qquad + \bigg( \frac{Y_1(ms^\alpha_{12})}{ms^\alpha_{12}} - \frac{Y_1(m\tau_{12})}{m\tau_{12}} \bigg) \cos\omega\tau_{12} \bigg] \,.
\end{align}
The divergence as $\tau_{12} \rightarrow 0$ in Eq.~\eqref{eq:besselrate} is independent of $\alpha$ and cancels. Thus, Eq.~\eqref{eq:besselratesubtracted} is independent of $\Lambda$ as $\tfrac{1}{\Lambda} \rightarrow 0$. Subtracting the inertial rate also gives an intuitive interpretation of the expression: it is the transition rate \textit{due to the detector's acceleration}.

If we are not fully inclusive over the final state of the radiation field and instead require that we remain in the Minkowski vacuum, then $E=\ket{0^{\phi}}\!\bra{0^{\phi}}\otimes\ket{2^D}\!\bra{2^D}$. In this case, we can quickly convince ourselves that the excitation probability --- now a single matrix element squared --- is zero, i.e.
\begin{equation}
\mathbb{P}(2;t)\ =\ \big|\!\braket{0^{\phi},2^D|U_{t,0}|0^{\phi},1^D}\!\big|^2\ =\ 0\,,
\end{equation}
since for this to be non-zero, $U_{t,0}$ would need an even number of field operators, $\phi(t,\mathbf{x}^D(t))$, and an odd number of monopole operators, $M^D(t)$. Since $H^{\rm int}(t)$ is linear in both $\phi(t,\mathbf{x}^D(t))$ and $M^D(t)$, this matrix element must therefore be zero. 

\subsection{Different Limits}

Considering Eq.~\eqref{eq:besselrate} in different limits can simplify the expression and act as a cross-check for numerical results, since all results derived in this section conform with the numerical results in Section \ref{sec:numericalresults}. First, the case of a massless scalar field ($m=0$) is considered. It is also shown that, for small acceleration $\alpha$, the subtracted rate scales as $\alpha^2$. At early times, the subtracted rate is independent of mass. At late times, the subtracted rate exhibits decaying oscillations as it tends to a constant, with the period and zeroes of the integrand of Eq.~\eqref{eq:largetimesubtractedrate} agreeing with the period and extrema of the numerical results.

\subsubsection{Massless limit \texorpdfstring{$(m \rightarrow 0)$}{(m -> 0)}}

When $m\rightarrow 0$, $\Delta^{R}_{12} \rightarrow 0$ and $\Delta^{H}_{12} \rightarrow \frac{-1}{2\pi^2 |s^\alpha_{12}|^2}$, which leaves
\begin{align}
\left.\mathbb{P}\right|_{m=0} &=\ \int^t_0 \D t_1\D t_2\frac{-|\mu|^2}{\gamma(t_1)\gamma(t_2)}\;\Theta_{12}\frac{\cos(\omega\tau_{12})}{2\pi^2(s^\alpha_{12})^2} \;; \\
\left.\frac{\partial \mathbb{P}}{\partial\tau} \right|_{m=0} &=\ -\frac{|\mu|^2\alpha^2}{8\pi^2}\int^\tau_{1/\Lambda} \D \tau_{12}\;\frac{\cos(\omega\tau_{12})}{\sinh^2\tfrac{1}{2}\alpha\tau_{12}} \;; \label{eq:acceleratedratemassless} \\
\left. \frac{\partial \mathbb{P}}{\partial \tau}\right|_{m=0} - \left.\frac{\partial \mathbb{P}}{\partial \tau}\right|_{m, \alpha = 0} \
&=\ -\frac{|\mu|^2}{8\pi^2}\int^\tau_{1/\Lambda} \D \tau_{12}\;\cos(\omega\tau_{12}) \bigg( \frac{\alpha^2}{\sinh^2\tfrac{1}{2}\alpha\tau_{12}} - \frac{4}{\tau_{12}^2} \bigg)\,.
\end{align}

\subsubsection{Small acceleration \texorpdfstring{$(\alpha \ll 1/\tau)$}{alpha << 1/tau}}

For $\alpha \ll 1/\tau$,
\begin{align}
    \frac{J_1 (m s^\alpha_{12})}{m s^\alpha_{12}} - \frac{J_1 (m \tau_{12})}{m \tau_{12}} &\rightarrow - \frac{\alpha^2 \tau^2}{24} J_2 (m \tau) + \mathcal{O}((\alpha\tau)^4)\,, \\
    \frac{Y_1 (m s^\alpha_{12})}{m s^\alpha_{12}} - \frac{Y_1 (m \tau_{12})}{m \tau_{12}} &\rightarrow - \frac{\alpha^2 \tau^2}{24} Y_2 (m \tau) + \mathcal{O}((\alpha\tau)^4)\,,
\end{align}
and Eq.~\eqref{eq:besselratesubtracted} becomes
\begin{align}
\label{eq:smallalpharate}
\frac{\partial \mathbb{P}}{\partial \tau} - \left.\frac{\partial \mathbb{P}}{\partial \tau}\right|_{\alpha = 0}
&=\ \frac{m^2 \alpha^2|\mu|^2}{96\pi} \int^\tau_{1/\Lambda} \D \tau_{12} \;\bigg[ J_2(m\tau_{12}) \sin\omega\tau_{12} + Y_2(m\tau_{12})\cos\omega\tau_{12} \bigg] \,.
\end{align}
Thus, for small $\alpha$, the rate scales as $\alpha^2$.

\subsubsection{Early times \texorpdfstring{$(\alpha\tau, 
 \, m\tau \rightarrow 0)$}{alpha*tau, m*tau -> 0}}

For $\alpha\tau, 
 \, m\tau \rightarrow 0$, we use
\begin{align}
    \frac{J_1 (m s^\alpha_{12})}{m s^\alpha_{12}} - \frac{J_1 (m \tau_{12})}{m \tau_{12}} &\rightarrow 0 \, ,\\
    \frac{Y_1 (m s^\alpha_{12})}{m s^\alpha_{12}} - \frac{Y_1 (m \tau_{12})}{m \tau_{12}} &\rightarrow \frac{\alpha^2}{6m^2 \pi} \, ,
\end{align}
and Eq.~\eqref{eq:besselratesubtracted} becomes
\begin{align}
\frac{\partial \mathbb{P}}{\partial \tau} - \left.\frac{\partial \mathbb{P}}{\partial \tau}\right|_{\alpha = 0}
&=\ \frac{\alpha^2 |\mu|^2}{24\pi^2} \tau \, .
\end{align}

\subsubsection{Late times \texorpdfstring{$(\tau \gg 1/\alpha)$}{tau >> 1/alpha}}

As $\tau_{12} \rightarrow \infty$, the integrand of Eq.~\eqref{eq:besselrate} goes to zero. This means that the rate tends to a constant as $\tau \rightarrow \infty$. This constant value is calculated in Section \ref{sec:momentumspace} (Eq.~\eqref{eq:acceleratedratelargetime}). However, as the rate tends to a constant, it also oscillates about the constant value. This is because, for large arguments,
\begin{align}
    J_1 (x) &\rightarrow \sqrt{\frac{2}{\pi x}} \cos (x - \frac{3\pi}{4})\,,\\
    Y_1 (x) &\rightarrow \sqrt{\frac{2}{\pi x}} \sin (x - \frac{3\pi}{4})\,,
\end{align}
such that (after taking $s^\alpha_{12} \gg \tau_{12}$),
\begin{align} \label{eq:largetimesubtractedrate}
\frac{\partial \mathbb{P}}{\partial \tau} - \left.\frac{\partial \mathbb{P}}{\partial \tau}\right|_{\alpha = 0} \
&=\  \text{constant } + |\mu|^2 \sqrt{\frac{m}{8\pi^3}} \int^\tau_{\tau_0} \D \tau_{12} \tau_{12}^{-3/2} \sin \biggl((m+\omega)\tau_{12} + \frac{\pi}{4} \biggr) \,,
\end{align}
where $\tau_0$ is a time large enough for the late-time limit to apply. 

\subsection{Momentum space}\label{sec:momentumspace}

The propagators in Eq.~\eqref{eq:Pmid} are Lorentz invariant and can be evaluated in any frame. Evaluating the momentum-space expressions for the propagators in the frame in which $x^0_{12} = s_{12}^{\alpha}$ and $\mathbf{x}_{12} =0$ gives
\begin{align}
    \Delta_{12} &= - \int \frac{\D^3 \mathbf{p}}{(2 \pi)^3} \frac{e^{i \mathbf{p \cdot x_{12}}}}{E_{\mathbf{p}}} \sin(E_{\mathbf{p}} x^0_{12}) = - \int \frac{\D^3 \mathbf{p}}{(2 \pi)^3} \frac{\sin(E_{\mathbf{p}} s_{12}^\alpha)}{E_{\mathbf{p}}}  \,,\\
    \Delta_{12}^H &= \int\frac{\D^3 \mathbf{p}}{(2 \pi)^3} \frac{e^{i \mathbf{p \cdot x_{12}}}}{E_{\mathbf{p}}} \cos(E_{\mathbf{p}} x^0_{12}) = \int\frac{\D^3 \mathbf{p}}{(2 \pi)^3} \frac{\cos(E_{\mathbf{p}} s_{12}^\alpha)}{E_{\mathbf{p}}} \,,
\end{align}
where
\begin{equation}
    E_{\mathbf{p}} = \sqrt{\mathbf{p}^2 + m^2} \,.
\end{equation}Inserting these expressions into Eq.~\eqref{eq:Pmid} gives
\begin{align}
\mathbb{P}(2;t) 	\ &=\ \frac{|\mu|^2}{8\pi^3}\int_0^{\tau} \D \tau_1\int_{0}^{\tau_1}\D \tau_2\;\int\D^2\mathbf{p}_{\perp}\int_{-\infty}^{\infty}\frac{\D p_x}{E_{\mathbf{p}}} \nonumber \\
& \qquad \Big[-\sin(E_{\mathbf{p}} s_{12}^\alpha)\sin(\omega \tau_{12})\:+\:\cos(E_{\mathbf{p}} s_{12}^\alpha)\cos(\omega \tau_{12})\Big] \nonumber \\
&=\ \frac{|\mu|^2}{8\pi^3}\int_0^{\tau} \D \tau_1\int_{0}^{\tau_1}\D \tau_2\;\int\D^2\mathbf{p}_{\perp}\int_{-\infty}^{\infty}\frac{\D p_x}{E_{\mathbf{p}}} \cos(E_{\mathbf{p}} s_{12}^\alpha + \omega \tau_{12}) \,.
\end{align}
Thus, the rate is
\begin{align}
\frac{\partial \mathbb{P}}{\partial \tau} &= \frac{|\mu|^2}{8\pi^3}\,\int_0^{\tau}\D \tau_{12}\int\D^2\mathbf{p}_{\perp}\int_{-\infty}^{\infty}\frac{\D p_x}{E_{\mathbf{p}}} \cos\left[\frac{2E_{\mathbf{p}}}{\alpha}\sinh(\alpha\tau_{12}/2)+\omega\tau_{12}\right] \nonumber \\
&= \frac{|\mu|^2}{16\pi^3}\,\int_{-\tau}^{\tau}\D \tau_{12}\int\D^2\mathbf{p}_{\perp}\int_{-\infty}^{\infty}\frac{\D p_x}{E_{\mathbf{p}}} \cos\left[\frac{2E_{\mathbf{p}}}{\alpha}\sinh(\alpha\tau_{12}/2)+\omega\tau_{12}\right]\,,
\label{eq:momentum_space_rate}
\end{align}
agreeing with the real part of Eq.~(3.12) of \cite{Crispino:2007eb}, which considers an initial state defined in the infinite past and a measurement taken in the infinite future (i.e., $\tau \rightarrow \infty$). In this limit, following \cite{Crispino:2007eb},
\begin{align}
\label{eq:acceleratedratelargetime}
\frac{\partial \mathbb{P}(\tau \rightarrow \infty)}{\partial \tau}
&= \frac{|\mu|^2}{2\pi^2 \alpha} e^{-\frac{\pi\omega}{\alpha}} \int_m^\infty \! {\textrm d}\nu \; \nu \bigg| K_{i\omega/\alpha} \bigg( \frac{\nu}{\alpha} \bigg) \bigg|^2 \, ,
\end{align}
where $\nu = \sqrt{\mathbf{p}_{\perp}^2 + m^2}$. In the limit $\alpha \rightarrow 0$, the integrand vanishes as $\frac{8\pi\alpha\nu}{\omega}e^{-\frac{\pi\omega}{\alpha}}\sin^2(\frac{\nu^2}{4\omega\alpha}+\ldots)$, which means an inertial detector does not undergo excitation in a vacuum at $\tau \rightarrow \infty$. As a result, at $\tau \to \infty$, the subtracted rate given by Eq.~\eqref{eq:besselratesubtracted} is equal to the unsubtracted rate given by Eq.~\eqref{eq:besselrate}.

Note that for an inertial path, $\mathbf{x}^D = (vt, 0, 0)$, the parameter $s^\alpha_{12}$ is replaced by
\begin{equation}
    s_{12} = \sqrt{(t_1 - t_2)^2 - v^2 \: (t_1 - t_2)^2} = \gamma^{-1} \; t_{12} = \tau_{12}\,,
\end{equation}
such that the transition rate becomes
\begin{align}
\label{eq:constant_v_rate}
\frac{\partial \mathbb{P}}{\partial \tau}
%&= \frac{|\mu|^2}{16\pi^3}\,\int_{-\tau}^{\tau}\D \tau_{12}\int\D^2\mathbf{p}_{\perp}\int_{-\infty}^{\infty}\frac{\D p_x}{E_{\mathbf{p}}} \cos\left[\gamma^{-1} \, E_{\mathbf{p}} t_{12} +\omega\tau_{12}\right] \nonumber\\ &
= \frac{|\mu|^2}{16\pi^3}\,\int_{-\tau}^{\tau}\D \tau_{12}\int\D^2\mathbf{p}_{\perp}\int_{-\infty}^{\infty}\frac{\D p_x}{E_{\mathbf{p}}} \cos\left[E_{\mathbf{p}} \tau_{12} +\omega\tau_{12}\right]\,.
\end{align}
As we will now show, we can start from this expression and re-derive Eq.~\eqref{eq:momentum_space_rate}. A derivation of the Unruh effect along these lines (for $m=0$) appears in \cite{Alsing:2004ig}. 

For the uniformly accelerated trajectory, the modes are subject to a characteristic, time-dependent Doppler shift, such that
\begin{equation}
    E_{\mathbf{p}}^{\prime}(\tau)=E_{\mathbf{p}}\cosh(\alpha\tau)-p_x \sinh(\alpha\tau)\,,
\end{equation}
reducing to
\begin{equation}
    E_{\mathbf{p}}^{\prime}(\tau)=E_{\mathbf{p}}e^{\mp\alpha\tau}\,,\qquad p_x\gtrless 0\,,
\end{equation}
in the massless limit and in one spatial dimension, as used in \cite{Alsing:2004ig}. To account for this, we can proceed from Eq.~\eqref{eq:constant_v_rate} by replacing
\begin{align}
    E_{\mathbf{p}}\tau_{12}\longrightarrow \int_{\tau_2}^{\tau_1}{\rm d}{\tau'}\,E_{\mathbf{p}}^{\prime}(\tau')&=\frac{E_{\mathbf{p}}}{\alpha}\left[\sinh(\alpha\tau_1)-\sinh(\alpha\tau_2)\right]-\frac{p_x}{\alpha}\left[\cosh(\alpha\tau_1)-\cosh(\alpha\tau_2)\right]\nonumber\\&=\frac{2}{\alpha}\sinh(\alpha\tau_{12}/2)\left[E_{\mathbf{p}}\cosh(\alpha\bar{\tau})-p_x\sinh(\alpha\bar{\tau})\right]\,,
\end{align}
where $\bar{\tau}=(\tau_1+\tau_2)/2$. The transition rate should not depend on $\bar{\tau}$. To see this, we boost to the instantaneous rest frame of the modes via the transformations
\begin{subequations}
\begin{align}
    E_{\mathbf{p}}^{\prime\prime}=E_{\mathbf{p}}\cosh(\alpha\bar{\tau})-p_x\sinh(\alpha\bar{\tau})\,,\\
    p_x^{\prime\prime}=p_x\cosh(\alpha\bar{\tau})-E_{\mathbf{p}}\sinh(\alpha\bar{\tau})\,.
\end{align}
\end{subequations}
The measure transforms as ${\rm d}p_x/E_{\mathbf{p}}={\rm d}p_x^{\prime\prime}/E_{\mathbf{p}}^{\prime\prime}$, and we recover Eq.~\eqref{eq:momentum_space_rate} after relabelling the integration variables.

\subsection{Transients}

The integral over $\tau_{12}\in[-\tau,\tau]$ in Eq.~\eqref{eq:momentum_space_rate} can be expressed as an integral over the whole real line by inserting a top-hat distribution of width $2\tau$, centred on the origin. Replacing the latter with its Fourier transform, we can write
\begin{align}
\frac{\partial \mathbb{P}}{\partial \tau} &
= \frac{|\mu|^2}{16\pi^3}\,\mathrm{Re}\,\int_{-\infty}^{\infty}\D \tau_{12}\int_{-\infty}^{\infty}\frac{{\rm d}k}{\pi}\,\frac{\sin(k\tau)}{k}\int\D^2\mathbf{p}_{\perp}\nonumber\\&\phantom{=}\times\int_{-\infty}^{\infty}\frac{\D p_x}{E_{\mathbf{p}}} \exp\left\{i\left[\frac{2E_{\mathbf{p}}}{\alpha}\sinh(\alpha\tau_{12}/2)+(\omega-k)\tau_{12}\right]\right\}\nonumber\\ &=\frac{|\mu|^2}{16\pi^3}\,\int_{-\infty}^{\infty}\D \tau_{12}\int_{-\infty}^{\infty}\frac{{\rm d}k}{\pi}\,\frac{\sin(k\tau)}{k}\int\D^2\mathbf{p}_{\perp}\int_{-\infty}^{\infty}\frac{\D p_x}{E_{\mathbf{p}}} \cos\left[\frac{2E_{\mathbf{p}}}{\alpha}\sinh(\alpha\tau_{12}/2)+(\omega-k)\tau_{12}\right]\,.
\end{align}
Swapping the order of the $\tau_{12}$ and $k$ integrals, we recognise the integral from~\cite{Crispino:2007eb} with $\omega \to \omega-k$, such that we have
\begin{align}
\frac{\partial \mathbb{P}(\tau)}{\partial \tau}
&= \frac{|\mu|^2}{2\pi^2 \alpha}\int_{-\infty}^{\infty}\frac{{\rm d}k}{\pi}\,\frac{\sin(k\tau)}{k}\, e^{-\frac{\pi(\omega-k)}{\alpha}} \int_m^\infty \! {\textrm d}\nu \; \nu \bigg| K_{i(\omega-k)/\alpha} \bigg( \frac{\nu}{\alpha} \bigg) \bigg|^2 \, .
\end{align}
In the limit, $\tau\to \infty$, we have
\begin{equation}
    \lim_{\tau\to\infty}\frac{1}{\pi}\,\frac{\sin(k \tau)}{k} = \delta(k)\,,
\end{equation}
and we recover Eq.~\eqref{eq:acceleratedratelargetime}.

Thus, we see that the transients arise from a convolution with the Fourier transform of the top-hat distribution. It is as if we fixed the field configuration at $t= -\infty$ and discontinuously turned on the interaction at $t=0$. If instead we turned the interaction on smoothly using some switching function then the transients would arise from a convolution with the Fourier transform of this switching function. 

\section{Excitation rate in a Rindler thermal bath} \label{sec:rindlerbath}

Before examining the results of the previous section, we shall compute the corresponding quantities from the perspective of a Rindler observer confined to the right Rindler wedge. The transformation from Minkowski to Rindler coordinates, $(\tau,\xi,y,z)$, is
\begin{equation}
    t = \alpha^{-1}\, e^{\alpha\xi}\, \sinh \alpha \tau \,, \quad x = \alpha^{-1} \,e^{\alpha\xi}\, \cosh \alpha \tau \,, \quad y=y \,, \quad z=z \,.
\end{equation}
In these coordinates, the atom is stationary at $\xi = 0$ (recovering Eqs.~\eqref{eq:trajectory} and \eqref{eq:propertime}), and the Minkowski vacuum state is exactly equivalent to a thermal state of Rindler particles. Therefore, from a Rindler observer's perspective, the detector is stationary in a thermal bath of Rindler particles at temperature $T$. In other words, the thermal bath is in an effective gravitational field (in accordance with the equivalence principle), and hence is different to a free-falling (inertial) thermal bath, which is considered in Section~\ref{sec:minkowskibath}.

The calculation proceeds in an identical fashion up to Eq.~\eqref{eq:accelrate}. We may expand the field using creation and annihilation operators for Rindler particles, i.e., in the right Rindler wedge~\cite{Crispino:2007eb, Takagi1986}
\begin{align}
    \phi(x) = \int \D E_{\mathbf{p}} \D^2 \mathbf{p}_\perp [v_{E_{\mathbf{p}}\mathbf{p}_\perp}^R \hat{a}_{E_{\mathbf{p}} \mathbf{p}_\perp}^R + \rm{H.c.}] \, ,
\end{align}
where `H.c.' stands for Hermitian conjugate and
\begin{align}
    v_{E_{\mathbf{p}}\mathbf{p}_\perp}^R &= \Bigg[ \frac{\sinh (\pi E_{\mathbf{p}} / \alpha)}{4 \pi^4 \alpha}\Bigg]^{1/2} K_{iE_{\mathbf{p}}/\alpha} \Bigg[ \frac{\sqrt{\mathbf{p}_\perp^2 + m^2}}{\alpha e^{-\alpha \xi}} \Bigg] e^{-i E_{\mathbf{p}} \tau + i \mathbf{p}_{\perp} \cdot \mathbf{x}_{\perp}} \nonumber \\
    &= \Bigg[ \frac{\sinh (E_{\mathbf{p}} / 2 \, T)}{8 \pi^5 T}\Bigg]^{1/2} K_{iE_{\mathbf{p}}/2\pi T} \Bigg[ \frac{\sqrt{\mathbf{p}_\perp^2 + m^2}}{2\pi T} \Bigg] e^{-i E_{\mathbf{p}} \tau + i \mathbf{p}_{\perp} \cdot \mathbf{x}_{\perp}}\,,
\end{align}
where $T = \alpha/2\pi$ is the Unruh temperature measured by an observer at the Rindler coordinate $\xi = 0$, and $\mathbf{x}_{\perp} = (y, z)$. The Rindler operators $\hat{a}_{E_{\mathbf{p}} \mathbf{p}_\perp}^R$ and $\hat{a}_{E_{\mathbf{p}} \mathbf{p}_\perp}^{R \, \dagger}$ define the Fock space for Rindler particles, and give the Minkowski vacuum expectation values~\cite{Crispino:2007eb}
\begin{align}
    \braket{0_{\phi}^{M} | \hat{a}_{E_{\mathbf{p}} \mathbf{p}_\perp}^{R \, \dagger} \hat{a}_{E'_{\mathbf{p}} \mathbf{p}'_\perp}^R | 0_{\phi}^{M}} &= ( e^{ E_{\mathbf{p}}/T} - 1)^{-1} \, \delta(E_{\mathbf{p}} - E'_{\mathbf{p}}) \,\delta^2(\mathbf{p}_\perp - \mathbf{p}'_\perp) \nonumber
    \\
    &= n \: \delta(E_{\mathbf{p}} - E'_{\mathbf{p}}) \,\delta^2(\mathbf{p}_\perp - \mathbf{p}'_\perp) \,,
    \\
    \braket{0_{\phi}^{M} | \hat{a}_{E_{\mathbf{p}} \mathbf{p}_\perp}^R \hat{a}_{E_{\mathbf{p}} \mathbf{p}_\perp}^{R \, \dagger} | 0_{\phi}^{M}} &= ( 1 - e^{- E_{\mathbf{p}}/T})^{-1} \,\delta(E_{\mathbf{p}} - E'_{\mathbf{p}}) \,\delta^2(\mathbf{p}_\perp - \mathbf{p}'_\perp) \nonumber\\
    &= (n+1) \, \delta(E_{\mathbf{p}} - E'_{\mathbf{p}}) \, \delta^2(\mathbf{p}_\perp - \mathbf{p}'_\perp)\,,
\end{align}
where
\begin{align}
    n \equiv n(E_{\mathbf{p}}) = ( e^{E_{\mathbf{p}}/T} - 1)^{-1}
\end{align}
is the Bose-Einstein distribution for a thermal bath at temperature $T$. For a stationary trajectory in Rindler coordinates at $\xi=0$, the Pauli-Jordan and Hadamard functions can be expressed as
\begin{align}
    \Delta_{12}\ &\equiv\ \frac{1}{i} \braket{0_{\phi}^{M}|[\phi_1,\phi_2]|0_{\phi}^{M}} \nonumber \\
    &= -\int \D E_{\mathbf{p}} \D^2 \mathbf{p}_\perp \frac{\sinh (E_{\mathbf{p}} / 2 T)}{4 \pi^5 T} \Bigg| K_{iE_{\mathbf{p}}/2\pi T} \Bigg[ \frac{\sqrt{\mathbf{p}_\perp^2 + m^2}}{2\pi T} \Bigg] \Bigg|^2 \sin (E_{\mathbf{p}} \tau_{12})  \,, \\
    \Delta^{H}_{12}\ &\equiv\ \braket{0_{\phi}^{M}|\{\phi_1,\phi_2\}|0_{\phi}^{M}} \nonumber \\
    &= \int \D E_{\mathbf{p}} \D^2 \mathbf{p}_\perp \frac{\sinh (E_{\mathbf{p}} / 2 T)}{4 \pi^5 T} \Bigg| K_{iE_{\mathbf{p}}/2\pi T} \Bigg[ \frac{\sqrt{\mathbf{p}_\perp^2 + m^2}}{2\pi T} \Bigg] \Bigg|^2 \cos (E_{\mathbf{p}} \tau_{12}) \big( 2 n + 1 \big) \,.
\end{align}
The transition rate (Eq.~\eqref{eq:accelrate}) then becomes
\begin{align}
    \frac{\partial \mathbb{P}}{\partial \tau} \ &= \frac{|\mu|^2}{4\pi^5 T}\int_0^\tau\! {\textrm d}\tau_{12} \int_0^\infty \! {\textrm d}E_{\mathbf{p}} \D^2 \mathbf{p}_\perp
    \sinh \bigg(\frac{E}{2T}\bigg) \bigg| K_{iE_{\mathbf{p}}/2\pi T} \bigg( \frac{\sqrt{\mathbf{p}_{\perp}^2 + m^2}}{2\pi T} \bigg) \bigg|^2 \times \nonumber \\
    & \qquad \Big( - \sin ( E_{\mathbf{p}} \tau_{12}) \sin ( \omega \tau_{12}) +(2n+1) \cos ( E_{\mathbf{p}} \tau_{12}) \cos ( \omega \tau_{12}) \Big) \nonumber \\
    &= \frac{|\mu|^2}{2\pi^4 T}\int_0^\infty \! {\textrm d}E_{\mathbf{p}} \int_m^\infty{\textrm d}\nu\:  \nu 
    \sinh \bigg(\frac{E}{2T}\bigg) \bigg| K_{iE_{\mathbf{p}}/2\pi T} \bigg( \frac{\nu}{2\pi T} \bigg) \bigg|^2 \times \nonumber \\
    & \qquad \Bigg( (n+1) \frac{\sin \big( (E_{\mathbf{p}} + \omega) \, \tau \big)}{E_{\mathbf{p}} + \omega} + n \: \frac{\sin \big( (E_{\mathbf{p}} - \omega) \, \tau \big)}{E_{\mathbf{p}} - \omega}
    \Bigg) \nonumber \\
    &= \frac{|\mu|^2}{4\pi^4 T}\int_0^\infty \! {\textrm d}E_{\mathbf{p}} \int_m^\infty{\textrm d}\nu\:  \nu 
     \bigg| K_{iE_{\mathbf{p}}/2\pi T} \bigg( \frac{\nu}{2\pi T} \bigg) \bigg|^2 \times \nonumber \\
    & \qquad \Bigg( e^{E_{\mathbf{p}} / 2T} \frac{\sin \big( (E_{\mathbf{p}} + \omega) \, \tau \big)}{E_{\mathbf{p}} + \omega} + e^{-E_{\mathbf{p}} / 2T} \: \frac{\sin \big( (E_{\mathbf{p}} - \omega) \, \tau \big)}{E_{\mathbf{p}} - \omega}
    \Bigg)\,. \label{eq:rindlerthermalrate}
\end{align}
The term proportional to $(n+1)$ in Eq.~\eqref{eq:rindlerthermalrate} corresponds to the \textit{emission} rate from the detector to the Rindler thermal bath, and the term proportional to $n$ corresponds to the \textit{absorption} rate of the detector from the Rindler thermal bath. The $\nu$ integral can be performed  when $m=0$~\cite{GradRyz}:
\begin{align}
    \left. \frac{\partial \mathbb{P}}{\partial \tau} \right|_{m=0} &= \frac{|\mu|^2 T}{\pi^2}\int_0^\infty \! {\textrm d}E_{\mathbf{p}}  
    \sinh \bigg(\frac{E_{\mathbf{p}}}{2T}\bigg) \bigg| \Gamma\bigg( 1+\frac{iE_{\mathbf{p}}}{2\pi T} \bigg) \bigg|^2 \times \nonumber \\
    & \qquad \Bigg( (n+1) \frac{\sin \big( (E_{\mathbf{p}} + \omega) \, \tau \big)}{E_{\mathbf{p}} + \omega} + n \: \frac{\sin \big( (E_{\mathbf{p}} - \omega) \, \tau \big)}{E_{\mathbf{p}} - \omega}
    \Bigg) \nonumber \\
    &= \frac{|\mu|^2}{2\pi^2}\int_0^\infty \! {\textrm d}E_{\mathbf{p}}  \,
    E_{\mathbf{p}} \Bigg( (n+1) \frac{\sin \big( (E_{\mathbf{p}} + \omega) \, \tau \big)}{E_{\mathbf{p}} + \omega} + n \: \frac{\sin \big( (E_{\mathbf{p}} - \omega) \, \tau \big)}{E_{\mathbf{p}} - \omega} \Bigg) \,,
\end{align}
where $\Gamma$ is the Gamma function and we have used
\begin{equation}
    \bigg| \Gamma\bigg( 1+\frac{iE_{\mathbf{p}}}{2\pi T} \bigg) \bigg|^2  = \frac{ E_{\mathbf{p}}}{2 T \sinh \big(E_{\mathbf{p}} / 2T \big)} \;.
\end{equation}

In order to match the calculation in Section \ref{sec:accel}, the inertial transition rate must be subtracted from Eq.~\eqref{eq:rindlerthermalrate}. From the Rindler observer's perspective, this corresponds to the $T=0$ limit of the transition rate. This is simply the transition rate for an inertial observer in the Minkowski vacuum, which is calculated and expressed as an integral over energy in Appendix \ref{app:inertialprob}. Note that the inertial rate is \textit{not} the same as simply taking the part of the Rindler thermal rate which is not proportional to $n$; doing this would give the transition rate for an accelerating detector in the Rindler vacuum. Subtracting Eq.~\eqref{eq:inertialrate}, the Rindler rate becomes
\begin{align} \label{eq:rindlerthermalratesubtracted}
    \frac{\partial \mathbb{P}}{\partial \tau} - \left.\frac{\partial \mathbb{P}}{\partial \tau}\right|_{T=0}
    &= \frac{|\mu|^2}{4\pi^4 T}\int_0^\infty \! {\textrm d}E_{\mathbf{p}} \int_m^\infty{\textrm d}\nu\:  \nu 
     \bigg| K_{iE_{\mathbf{p}}/2\pi T} \bigg( \frac{\nu}{2\pi T} \bigg) \bigg|^2 \times \nonumber \\
    & \qquad \Bigg( e^{E_{\mathbf{p}} / 2T} \frac{\sin \big( (E_{\mathbf{p}} + \omega) \, \tau \big)}{E_{\mathbf{p}} + \omega} + e^{-E_{\mathbf{p}} / 2T} \: \frac{\sin \big( (E_{\mathbf{p}} - \omega) \, \tau \big)}{E_{\mathbf{p}} - \omega}
    \Bigg) \nonumber \\
    & \qquad - \frac{|\mu|^2}{2\pi^2}\int_m^{\infty}{\rm d}E\;\sqrt{E^2-m^2}\;\frac{\sin[(E+\omega)\tau]}{E+\omega}
    \;.
\end{align}
We have checked numerically that this is the same result as Eq.~\eqref{eq:besselratesubtracted} for all times. Eq.~\eqref{eq:rindlerthermalratesubtracted} can be shown to vanish when $T \to 0$, since
\begin{equation}
    \lim_{T\to 0} \int_m^\infty{\textrm d}\nu\:  \nu 
     \bigg| K_{iE_{\mathbf{p}}/2\pi T} \bigg( \frac{\nu}{2\pi T} \bigg) \bigg|^2 = e^{-E_{\mathbf{p}}/ 2T} \; 2 \pi^2 T \; \Theta (E_{\mathbf{p}} - m) \sqrt{E_{\mathbf{p}}^2 - m^2} \,,
\end{equation}
where $\Theta (x)$ is the Heaviside function. Taking the $\tau \rightarrow \infty$ limit of Eq.~\eqref{eq:rindlerthermalratesubtracted} gives the transition rate of a detector coupled to a massive scalar field after the transient effects have subsided. In this limit, Eq.~\eqref{eq:rindlerthermalratesubtracted} becomes,
\begin{align}
\label{eq:rindlerratelargetime}
\frac{\partial \mathbb{P}(\tau \rightarrow \infty)}{\partial \tau} - \left. \frac{\partial \mathbb{P}(\tau \rightarrow \infty)}{\partial \tau}\right|_{T=0}
&= \frac{|\mu|^2}{4\pi^3 T} e^{-\frac{\omega}{2T}} \int_m^\infty \! {\textrm d}\nu \; \nu \bigg| K_{i\omega/2\pi T} \bigg( \frac{\nu}{2\pi T} \bigg) \bigg|^2 \,,
\end{align}
since
\begin{equation}
\lim_{\tau\to \infty}\frac{\sin[(E_{\mathbf{p}} \pm\omega)\tau]}{E_{\mathbf{p}} \pm\omega}=\pi\delta(E_{\mathbf{p}} \pm\omega)\,,
\end{equation}
and only the delta function $\delta(E_{\mathbf{p}}-\omega)$ has support over the domain of the energy integrals. This corresponds to the emission rate vanishing at late times (once the transients have decayed), so the detector only absorbs Rindler particles and does not emit them. This is the same result as Eq.~\eqref{eq:acceleratedratelargetime}.

For a massless scalar field
\begin{align} \label{eq:rindlermasslessratesubtracted}
    \left. \frac{\partial \mathbb{P}}{\partial \tau} \right|_{m=0} - \left.\frac{\partial \mathbb{P}}{\partial \tau}\right|_{T,m=0}
    &= \frac{|\mu|^2}{2\pi^2}\int_0^\infty \! {\textrm d}E_{\mathbf{p}}  \,
    E_{\mathbf{p}} \, n \, \Bigg( \frac{\sin \big( (E_{\mathbf{p}} + \omega) \, \tau \big)}{E_{\mathbf{p}} + \omega} + \: \frac{\sin \big( (E_{\mathbf{p}} - \omega) \, \tau \big)}{E_{\mathbf{p}} - \omega} \Bigg)
    \,,
\end{align}
which, in the $\tau \to \infty$ limit, reduces to
\begin{align}
\left. \frac{\partial \mathbb{P}(\tau \to \infty)}{\partial \tau} \right|_{m=0} - \left.\frac{\partial \mathbb{P} (\tau \to \infty)}{\partial \tau}\right|_{m, T=0}
&=  \frac{|\mu|^2}{2\pi} \frac{ \omega}{e^{\beta\omega}-1} \, .
\end{align}
Coincidentally, this is the the expected transition rate for a detector in an inertial Bose-Einstein thermal bath of massless Minkowski particles, as will be shown in the next section.

\section{Excitation rate in a Minkowski thermal bath} \label{sec:minkowskibath}

We shall now calculate the response of a detector in a thermal bath of Minkowski particles and show that this is different to the Rindler case for $m \neq 0$. 
The zero-temperature, Minkowski Hadamard function is given by
\begin{align}
\label{eq:MinkHadamard}
    \left. \Delta^{H}_{12} \right|_{T=0} = \left.\Delta^>_{12}\right|_{T=0} + \left.\Delta^<_{12}\right|_{T=0} = \int \frac{\D^4 p}{(2\pi)^3} \, e^{-ip_\mu x_{12}^\mu} \, \delta(p^2 - m^2) \,.
\end{align}
In equilibrium with a thermal bath at temperature, $T$, the Wightman functions are
\begin{align}
    \Delta^>_{12}(T) &= \int \frac{\D^4 p}{(2\pi)^3} \Big[ \Theta(p^0) \,(1 + n) + \Theta(-p^0) \, n \Big] e^{-ip_\mu x_{12}^\mu} \,\delta(p^2 - m^2)\,, \\
    \Delta^<_{12}(T) &= \int \frac{\D^4 p}{(2\pi)^3} \Big[ \Theta(-p^0) \,(1 + n) + \Theta(p^0) \, n \Big] e^{-ip_\mu x_{12}^\mu} \,\delta(p^2 - m^2) \,.
\end{align}
where $n \equiv n(|p^0|) = (\exp(|p^0|/T) - 1)^{-1}$. The thermal Minkowski Hadamard function is therefore
\begin{align}
    \Delta^{H}_{12}(T) &= \int \frac{\D^4 p}{(2\pi)^3} \Big[ \Theta(p^0) + \Theta(-p^0) \Big] \big( 1 + 2n \big) e^{-ip_\mu x_{12}^\mu} \,\delta(p^2 - m^2) 
    \\
    &= \int \frac{\D^4 p}{(2\pi)^3} \big( 1 + 2n \big) e^{-ip_\mu x_{12}^\mu} \,\delta(p^2 - m^2)\,.
\end{align}
Comparing with Eq.~\eqref{eq:MinkHadamard}, we see that the thermal piece of the Hadamard function is simply the zero-temperature piece multiplied by $2n$. Since the detector is static, $\mathbf{x}_{12} = 0$ (as long as the detector is inertial, we can boost to this frame due to its time-like trajectory). Therefore, the thermal part of $\Delta^{H}_{12}$ takes the form,
\begin{align}
\Delta^{H}_{12}&\supset\int\!\frac{{\rm d}^4p}{(2\pi)^3}\;e^{-ip^0t_{12}} \: (2 n) \:\delta(p^2-m^2)=\frac{1}{\pi^2}\int_{m}^{\infty}{\rm d}E\;\sqrt{E^2-m^2}\:\, n \,\cos(Et_{12})\,.
\end{align}
The retarded propagator does not pick up a thermal part (i.e., it does not have any temperature dependence, since terms proportional to $n$ cancel when taking the difference of positive and negative Wightman functions). The thermal contribution to the excitation rate is thus
\begin{align}
\label{eq:thermalrate}
    \frac{\partial \mathbb{P}}{\partial t} - \left.\frac{\partial \mathbb{P}}{\partial t}\right|_{\alpha=0}
    &= \ \frac{|\mu|^2}{\pi^2} \int_0^{t} \D t' \int_m^\infty \D E \sqrt{E^2-m^2} \:\, n \, \cos\omega t'\cos Et'  \nonumber \\
    &= \frac{|\mu|^2}{2\pi^2}\int_m^{\infty}{\rm d}E\;\sqrt{E^2-m^2}\:\, n \, \left\{\frac{\sin[(E-\omega)t]}{E-\omega}+\frac{\sin[(E+\omega)t]}{E+\omega}\right\}\,.
\end{align}
This expression is not equal to Eq.~\eqref{eq:rindlerthermalrate}. However, with $m=0$, this expression is exactly the same as Eq.~\eqref{eq:rindlermasslessratesubtracted}. This means that the response of a \textit{monopole detector} coupled to a \textit{massless} scalar field is insensitive to the difference between inertial and accelerating thermal distributions. This misleading example may lead one to the conclusion that an accelerated detector responds identically to an inertial detector in an ordinary (Minkowski) Bose gas at finite temperature. This is not the statement of the Unruh effect. It is clearly not true for a massive scalar field, nor is it true for vector fields or other detector models~\cite{PhysRevD.29.1047, Crispino:2007eb, Fulling:2014}.

In the limit $t\to\infty$, we obtain
\begin{align}
\frac{\partial \mathbb{P}}{\partial t}&\supset \frac{|\mu|^2}{2\pi}\int_m^{\infty}{\rm d}E\;\sqrt{E^2-m^2}\,\frac{1}{e^{\beta E}-1}\,\Big[\delta(E-\omega)+\delta(E+\omega)\Big]\,.
\end{align}
Only the first delta function has support, and only when $\omega \geq m$, so we arrive at the result
\begin{align}
\label{eq:boseeinsteinrate}
\frac{\partial \mathbb{P}}{\partial t}&\supset \frac{|\mu|^2}{2\pi}\frac{\sqrt{\omega^2-m^2}}{e^{\beta \omega}-1} \Theta(\omega - m)\,,
\end{align}
agreeing with Eq.~(3.73) of \cite{Birrell:1982ix}. Thus, a detector in a thermal bath of Minkowski particles requires $\omega > m$. This is not the case for the Rindler thermal bath, and this highlights a crucial physical difference. For the Rindler thermal bath, as $\tau \to \infty$, it remains true that the energy, $E$, of the absorbed Rindler particle must equal the detector's energy gap, $\omega$, but the transition rate is non-zero even if the energy gap is less than the mass of the field ($\omega < m$). This is because $E \geq m$ is a flat spacetime constraint, and a general field quantization does not lead to a simple dispersion relation relating a particle's energy to its mass. This difference can be seen when comparing Figures~\ref{fig:largetimerates} and~\ref{fig:minklargetimerates}. Further discussion of Rindler particles with energy $E < m$ is given in Section III.A.3 of \cite{Crispino:2007eb}.

\section{Numerical Results} \label{sec:numericalresults}

The transition rate for a uniformly accelerating detector, with the inertial rate subtracted, is given by the identical Eqs.~\eqref{eq:besselratesubtracted} and~\eqref{eq:rindlerthermalratesubtracted}. The explicit dependence of the transition rate on the detector's proper time is shown in Figure~\ref{fig:manyalphamanymass}. Specifying the state at $\tau = 0$ causes transients and the frequency of these transients is dependent on $m/\omega$ and independent of $\alpha/\omega$. As $\omega\tau \to \infty$, the transients decay and the rate tends to a constant value. We can also observe that the transients subside more rapidly for larger accelerations.  

The constant, late-time value is given by Eq.~\eqref{eq:acceleratedratelargetime} (and Eq.~\eqref{eq:rindlerratelargetime}). The dependence of this `equilibrium' rate on temperature (and hence acceleration via $T = \alpha / 2 \pi$) and mass is shown in Figure \ref{fig:largetimerates}. The values of $\alpha/\omega$ and $m/\omega$ are chosen so as to scan a wide range of dimensionless ratios. Figure~\ref{fig:largetimeratevsalpha} shows that the larger the detector's acceleration, the larger the transition rate at $\tau \to \infty$. It also shows that for a scalar field with larger mass, a larger acceleration is required for the detector to `switch on', with the transition rate becoming non-negligible at $T \sim m/4\pi$ ($\alpha \sim m/2$). This is superseded by another requirement: the detector `switches on' at $T/\omega \sim 1/2\pi$ ($\alpha \sim \omega$). This leads to the sensible conclusion that the detector's transition rate begins to increase when its acceleration is above its energy gap. At large accelerations, the gradient becomes independent of the mass, meaning that the `sensitivity' of the transition rate to acceleration (defined as $\D^2 \mathbb{P} / \D \tau \D \alpha$) is independent of mass. Figure~\ref{fig:largetimeratevsmass} shows how the transition rate at $\tau \to \infty$ depends on the mass of the scalar field. As the mass increases, the transition rate tends to zero. However, it remains non-zero above $m/\omega=1$, which reflects the fact that an accelerating detector can absorb quanta of larger mass than its energy gap.

To highlight the difference between a Rindler thermal bath and a Minkowski thermal bath, the transition rate at $\tau\to\infty$ for a detector in a Minkowski thermal bath is plotted in Figure \ref{fig:minklargetimerates}. Figure~\ref{fig:minkowskiratevsalpha} shows that, like the Rindler bath case, the transition rate for the Minkowski thermal bath also `switches on' at $T/\omega \sim 1/2\pi$ ($\alpha \sim \omega$), but there is no longer any requirement that $T\gtrsim m/4\pi$ ($\alpha \gtrsim m/2$) and the gradient of the rate (the sensitivity) is dependent on the mass even at large accelerations. Also, the transition rate at $\tau \to \infty$ is zero for $m \geq \omega$, regardless of the temperature. This is due to the flat-spacetime constraint $E\geq m$, which means the detector cannot absorb a particle of mass larger than its energy gap. Figures~\ref{fig:largetimeratevsmass} and~\ref{fig:minkowskiratevsmass} illustrate that, if $m=0$, the $\tau \to \infty$ transition rate of a detector in a Rindler thermal bath is identical to that of a detector in a Minkowksi thermal bath. The transition rate then differs as $m/\omega$ is increased. This is true for all times, not just $\tau \to \infty$, as explained in Section~\ref{sec:minkowskibath}.

\begin{figure}[!htb]
\centering
\begin{subfigure}{.5\textwidth}
  \centering
  \includegraphics[width=\linewidth]{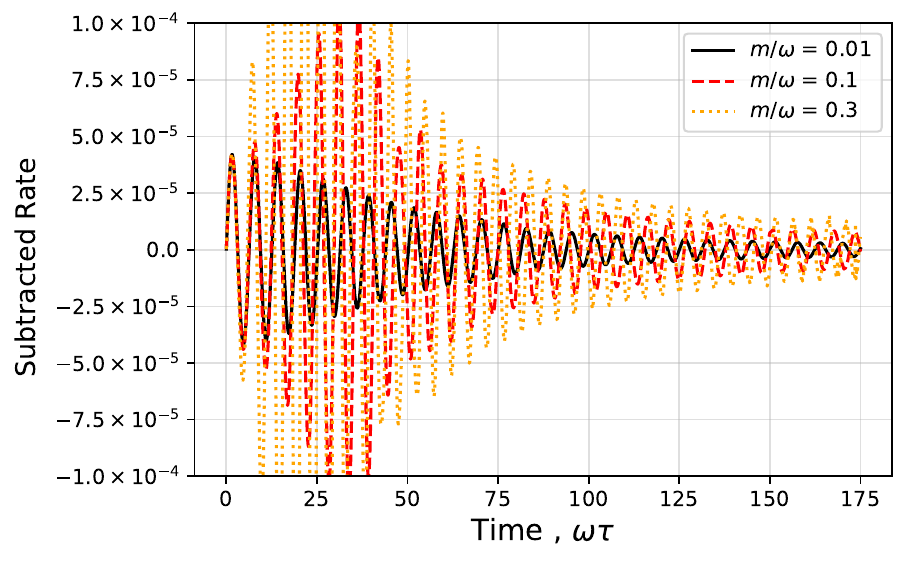}
  \caption{$\alpha/\omega = 0.1$}
  \label{fig:largerangealpha01bessel}
\end{subfigure}%
\begin{subfigure}{.5\textwidth}
  \centering
  \includegraphics[width=\linewidth]{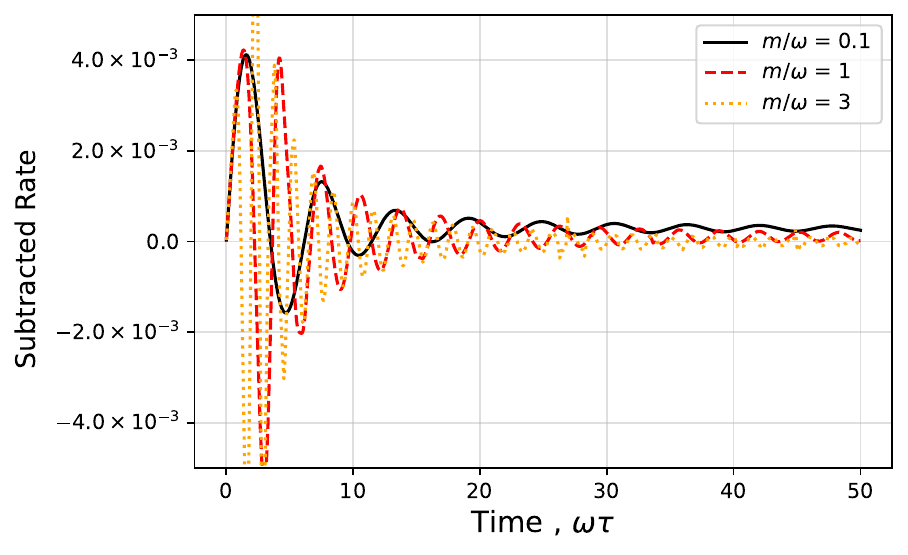}
  \caption{$\alpha/\omega = 1$}
  \label{fig:largerangealpha1bessel}
\end{subfigure}
\\
\begin{subfigure}{.5\textwidth}
  \centering
  \includegraphics[width=\linewidth]{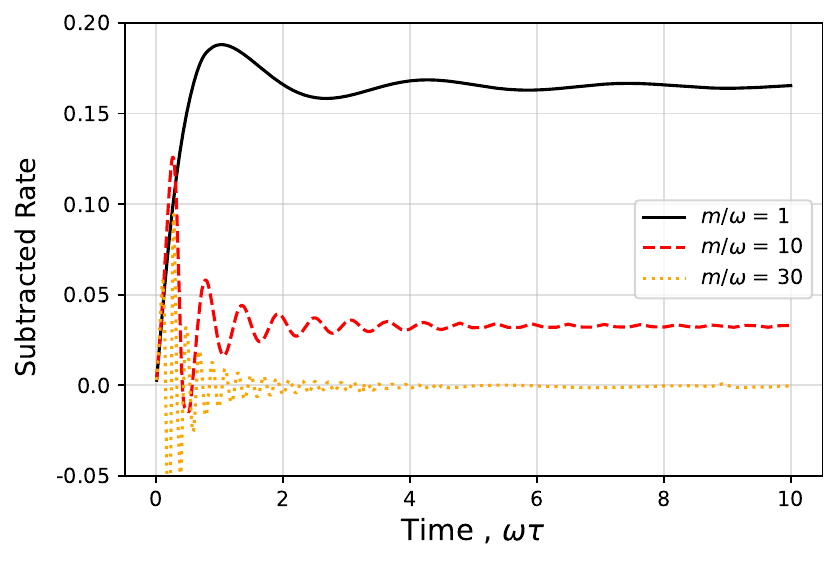}
  \caption{$\alpha/\omega = 10$}
  \label{fig:largerangealpha10bessel}
\end{subfigure}%
\begin{subfigure}{.5\textwidth}
  \centering
  \includegraphics[width=\linewidth]{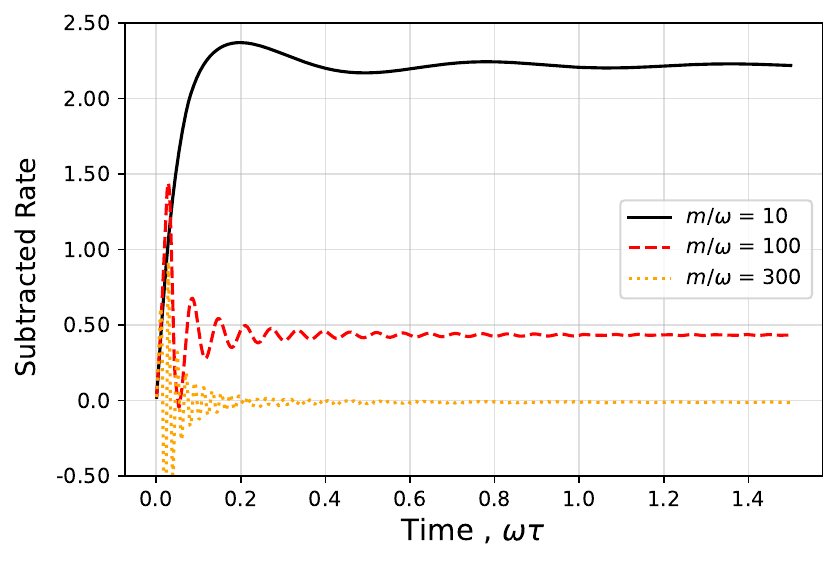}
  \caption{$\alpha/\omega = 100$}
  \label{fig:largerangealpha100bessel}
\end{subfigure}
\caption{The transition rate for an accelerated detector against time, with the inertial transition rate subtracted (given by Eq.~\eqref{eq:besselratesubtracted}), 
i.e., $\frac{1}{\omega|\mu|^2}\!\left(\frac{\D\mathbb{P}}{\D\tau} - \left.\frac{\D\mathbb{P}}{\D\tau}\right|_{\alpha=0}\right)\ $.
Each plot is for a given acceleration, with three different values of the mass of the scalar field. The transition rate exhibits transient effects, but at late times tends to a constant value.}
\label{fig:manyalphamanymass}
\end{figure}

\begin{figure}[!htb]
\centering
\begin{subfigure}{.5\textwidth}
  \centering
  \includegraphics[width=\linewidth]{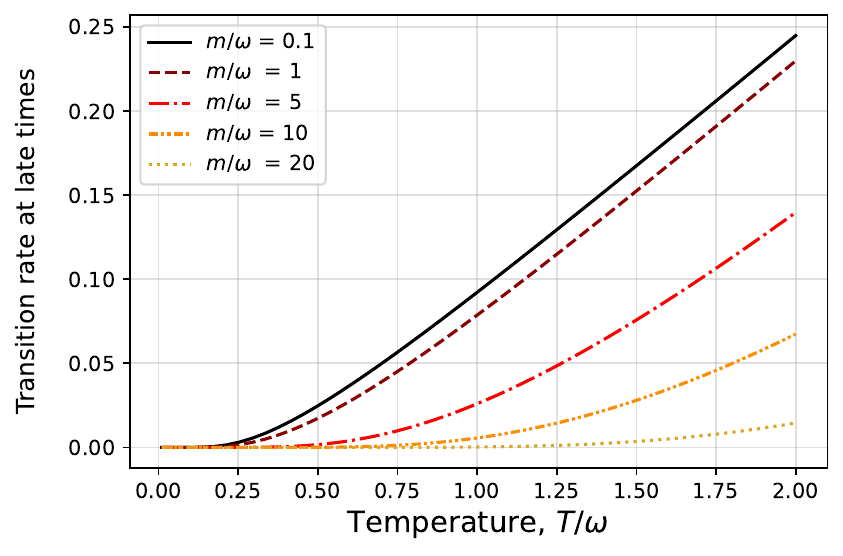}
  \caption{}
  \label{fig:largetimeratevsalpha}
\end{subfigure}%
\begin{subfigure}{.5\textwidth}
  \centering
  \includegraphics[width=\linewidth]{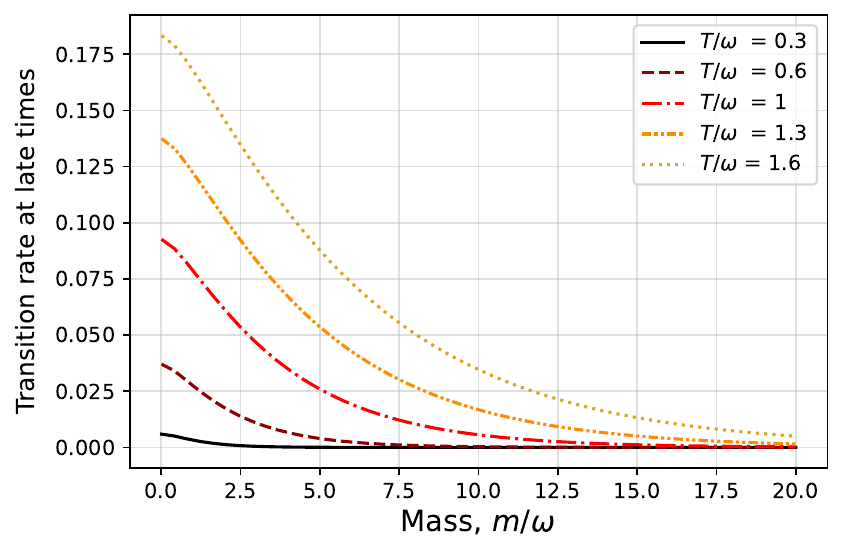}
  \caption{}
  \label{fig:largetimeratevsmass}
\end{subfigure}
\caption{The transition rate for an {\it accelerated} detector tends to a constant at late times, given by the identical Eqs.~\eqref{eq:acceleratedratelargetime} and~\eqref{eq:rindlerratelargetime}. This constant depends on the acceleration (left) and the mass of the scalar field (right) as plotted here.}
\label{fig:largetimerates}
\end{figure}

\begin{figure}[!htb]
\centering
\begin{subfigure}{.5\textwidth}
  \centering
  \includegraphics[width=\linewidth]{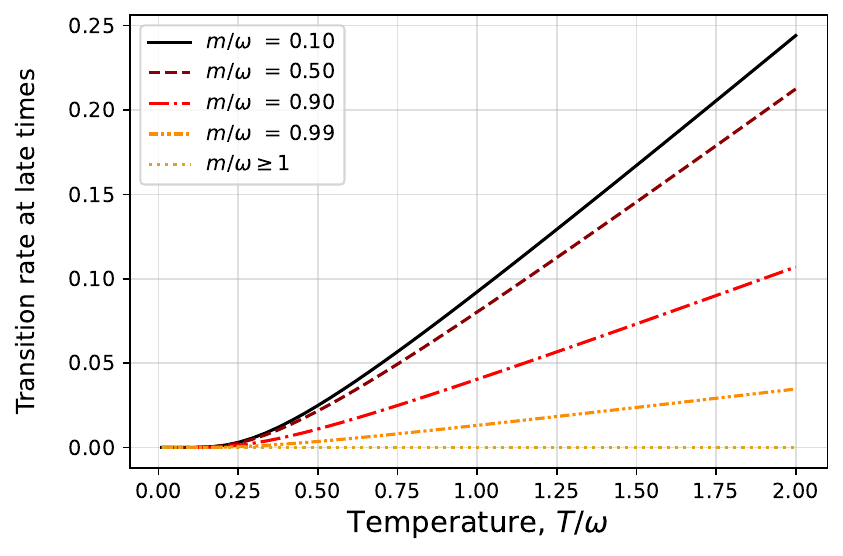}
  \caption{}
  \label{fig:minkowskiratevsalpha}
\end{subfigure}%
\begin{subfigure}{.5\textwidth}
  \centering
  \includegraphics[width=\linewidth]{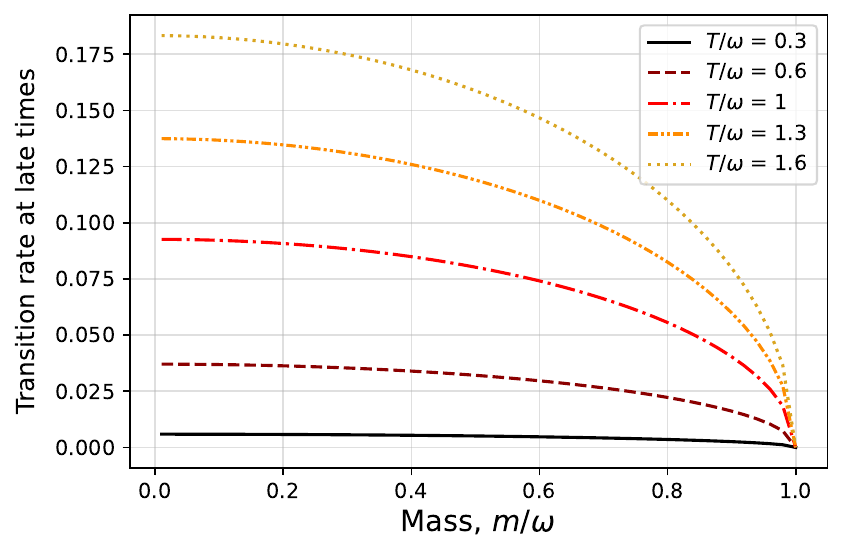}
  \caption{}
  \label{fig:minkowskiratevsmass}
\end{subfigure}
\caption{The transition rate for an {\it inertial} detector in a Minkowski thermal bath tends to a constant at late times, given by Eq.~\eqref{eq:boseeinsteinrate}. This constant depends on the acceleration (left) and the mass of the scalar field (right) as plotted here. }
\label{fig:minklargetimerates}
\end{figure}

\section{Conclusion} \label{sec:conc}

We have employed a probabilistic method to calculate the first-order transition rate of a uniformly accelerated Unruh-DeWitt monopole detector from the ground state to the excited state, with the inertial rate subtracted (Eq.~\eqref{eq:besselratesubtracted}). The transition rate has been expressed as a sum of two terms; one is proportional to $\Delta^R$ and independent of the initial state, and one is proportional to $\Delta^H$ and encapsulates all initial state dependence. Eq.~\eqref{eq:rindlerthermalrate} is the same transition rate, calculated from the perspective of a Rindler (accelerating) observer, who describes the detector as stationary in Rindler coordinates in a thermal bath of Rindler particles. The two expressions are equal at all times, including the transient effects which arise due to specifying the field to initially be in the Minkowski vacuum state. This is due to the Unruh effect: an observer accelerating through the Minkowski vacuum experiences a thermal bath of Rindler particles. Eq.~\eqref{eq:thermalrate} is the corresponding transition rate for an inertial detector in a `real' (Minkowski) thermal bath. This rate is different and is unrelated to the Unruh effect. It is only coincidentally equal for a massless scalar field. The Unruh effect has also been presented as the result of a time-dependent Doppler shift of the field modes. The numerical results are new and highlight the dependence of the transients on the mass of the scalar field, the acceleration and the energy gap of the detector. The late-time behaviour of the transition rate has been explored numerically and compared to the transition rate for an inertial detector in the Minkowski thermal bath.

The probability-level framework presented here can be utilised to study the response of an accelerated detector with a different model of the detector, a smooth switching function for the interaction (resulting in different transients), or different background spacetimes. 
It also has the advantage of being able to treat mixed states and (semi-)inclusive observables, potentially simplifying calculations that are more complex at the amplitude level. We also hope that our approach may be of use to studies of relativistic quantum information, in which it is natural to work with mixed states.

\begin{acknowledgments}

PM would like to thank Florian Niedermann for discussions in the early stages of this work. This work was supported by the Science and Technology Facilities Council (STFC) [Grant No.~ST/X00077X/1], a United Kingdom Research and Innovation (UKRI) Future Leaders Fellowship [Grant No.~MR/V021974/2], and a Leverhulme Trust Research Leadership Award [Grant No.~RL-2016-028]. RJ would like to thank Callsign Ltd for their financial support. 

\end{acknowledgments}

\section*{Data Access Statement}

No data were created or analysed in this study.

\appendix

\section{Inertial Excitation Rate in Minkowski Vacuum} \label{app:inertialprob}
Here we consider the excitation probability of an inertial detector in the Minkowski vacuum. For a time-like interval, one can always boost to a frame in which $\mathbf{x}_1 - \mathbf{x}_2 = 0$, such that,
\begin{subequations}
\begin{align}
\left.\Delta^{R}_{12}\right|_{\alpha=0}&=\int\!\frac{{\rm d}^4p}{(2\pi)^4}\;\frac{e^{-ip^0t_{12}}}{(p^0+i\epsilon)^2-E_{\mathbf{p}}^2}=-\frac{1}{2\pi^2}\int_{m}^{\infty}{\rm d}E\;\sqrt{E^2-m^2}\,\sin(Et_{12}),\\
\left.\Delta^{H}_{12}\right|_{\alpha=0}&=\int\!\frac{{\rm d}^4p}{(2\pi)^4}\;e^{-ip^0t_{12}}2\pi\delta(p^2-m^2)=\frac{1}{2\pi^2}\int_{m}^{\infty}{\rm d}E\;\sqrt{E^2-m^2}\,\cos(Et_{12}).
\end{align}
\end{subequations}
Using Eq.~\eqref{eq:Pmid}, we have
\begin{align}
\left.\mathbb{P}(2;t)\right|_{\alpha=0}&=\frac{|\mu|^2}{2\pi^2}\int_{0}^{t}{\rm d}t_1\int_{0}^{t_1}{\rm d}t_2\;\int_m^{\infty}{\rm d}E\;\sqrt{E^2-m^2}\left[\cos(\omega t_{12})\cos(E t_{12})-\sin(\omega t_{12})\sin(E t_{12})\right]\nonumber\\&=\frac{|\mu|^2}{2\pi^2}\int_{0}^{t}{\rm d}t_1\int_{0}^{t_1}{\rm d}t_2\;\int_m^{\infty}{\rm d}E\;\sqrt{E^2-m^2}\,\cos[(E+\omega)t_{12}].
\end{align}
Performing the time integrals, we have
\begin{equation}
\left.\mathbb{P}(2;t)\right|_{\alpha=0}=\frac{|\mu|^2}{2\pi^2}\int_m^{\infty}{\rm d}E\;\sqrt{E^2-m^2}\;\frac{1-\cos[(E+\omega)t]}{(E+\omega)^2},
\end{equation}
such that the transition rate is
\begin{equation} \label{eq:inertialrate}
\left.\frac{\partial \mathbb{P}(2;t)}{\partial t}\right|_{\alpha=0}=\frac{|\mu|^2}{2\pi^2}\int_m^{\infty}{\rm d}E\;\sqrt{E^2-m^2}\;\frac{\sin[(E+\omega)t]}{E+\omega}.
\end{equation}
This transition rate exhibits an ultraviolet divergence due to the treatment of the detector as point-like. By considering the difference of two transition rates (e.g., accelerated rate minus inertial rate), we subtract this divergence.

\bibliography{references}

\end{document}